\documentclass[aps,prl,showpacs,twocolumn,lineno,groupedaddress]{revtex4}  
\usepackage{graphicx}  
\usepackage{dcolumn}   
\usepackage{bm}        
\usepackage{amssymb}   

\begin{document}



\def\slashchar#1{\setbox0=\hbox{$#1$}           
   \dimen0=\wd0                                 
   \setbox1=\hbox{/} \dimen1=\wd1               
   \ifdim\dimen0>\dimen1                        
      \rlap{\hbox to \dimen0{\hfil/\hfil}}      
      #1                                        
   \else                                        
      \rlap{\hbox to \dimen1{\hfil$#1$\hfil}}   
      /                                         
   \fi}  
                                            
\def\etmiss{\slashchar{E}_T}

\hspace{5.2in} \mbox{FERMILAB-PUB-04-389-E}

\title{Search for first-generation scalar leptoquarks in
$\bm{p \bar{p}}$ collisions at $\sqrt{s}$=1.96 TeV\\ }
%
\author{                                                                      
V.M.~Abazov,$^{34}$                                                           
B.~Abbott,$^{71}$                                                             
M.~Abolins,$^{62}$                                                            
B.S.~Acharya,$^{28}$                                                          
M.~Adams,$^{49}$                                                              
T.~Adams,$^{47}$                                                              
M.~Agelou,$^{17}$                                                             
J.-L.~Agram,$^{18}$                                                           
S.H.~Ahn,$^{30}$                                                              
M.~Ahsan,$^{56}$
% ...
\\                                                                            
\vskip 0.30cm                                                                 
\centerline{(D\O\ Collaboration)}                                             
\vskip 0.30cm                                                                 
}                                                                             
\address{                                                                     
\centerline{$^{1}$Universidad de Buenos Aires, Buenos Aires, Argentina}       
\centerline{$^{2}$LAFEX, Centro Brasileiro de Pesquisas F{\'\i}sicas,         
                  Rio de Janeiro, Brazil}                                     
\centerline{$^{3}$Universidade do Estado do Rio de Janeiro,                   
                  Rio de Janeiro, Brazil}                                     
\centerline{$^{4}$Instituto de F\'{\i}sica Te\'orica, Universidade            
                  Estadual Paulista, S\~ao Paulo, Brazil}                     
\centerline{$^{5}$University of Alberta, Edmonton, Alberta, Canada,           
               Simon Fraser University, Burnaby, British Columbia, Canada,}   
\centerline{York University, Toronto, Ontario, Canada, and                    
         McGill University, Montreal, Quebec, Canada}                         
\centerline{$^{6}$Institute of High Energy Physics, Beijing,                  
                  People's Republic of China}                                 
\centerline{$^{7}$Universidad de los Andes, Bogot\'{a}, Colombia}             
\centerline{$^{8}$Center for Particle Physics, Charles University,            
                  Prague, Czech Republic}                                     
\centerline{$^{9}$Czech Technical University, Prague, Czech Republic}         
\centerline{$^{10}$Institute of Physics, Academy of Sciences, Center          
                  for Particle Physics, Prague, Czech Republic}               
\centerline{$^{11}$Universidad San Francisco de Quito, Quito, Ecuador}        
\centerline{$^{12}$Laboratoire de Physique Corpusculaire, IN2P3-CNRS,         
                 Universit\'e Blaise Pascal, Clermont-Ferrand, France}        
\centerline{$^{13}$Laboratoire de Physique Subatomique et de Cosmologie,      
                  IN2P3-CNRS, Universite de Grenoble 1, Grenoble, France}     
\centerline{$^{14}$CPPM, IN2P3-CNRS, Universit\'e de la M\'editerran\'ee,     
                  Marseille, France}                                          
\centerline{$^{15}$Laboratoire de l'Acc\'el\'erateur Lin\'eaire,              
                  IN2P3-CNRS, Orsay, France}                                  
\centerline{$^{16}$LPNHE, IN2P3-CNRS, Universit\'es Paris VI and VII,         
                  Paris, France}                                              
\centerline{$^{17}$DAPNIA/Service de Physique des Particules, CEA, Saclay,    
                  France}                                                     
\centerline{$^{18}$IReS, IN2P3-CNRS, Universit\'e Louis Pasteur, Strasbourg,  
                France, and Universit\'e de Haute Alsace, Mulhouse, France}   
\centerline{$^{19}$Institut de Physique Nucl\'eaire de Lyon, IN2P3-CNRS,      
                   Universit\'e Claude Bernard, Villeurbanne, France}         
\centerline{$^{20}$III. Physikalisches Institut A, RWTH Aachen,               
                   Aachen, Germany}                                           
\centerline{$^{21}$Physikalisches Institut, Universit{\"a}t Bonn,             
                  Bonn, Germany}                                              
\centerline{$^{22}$Physikalisches Institut, Universit{\"a}t Freiburg,         
                  Freiburg, Germany}                                          
\centerline{$^{23}$Institut f{\"u}r Physik, Universit{\"a}t Mainz,            
                  Mainz, Germany}                                             
\centerline{$^{24}$Ludwig-Maximilians-Universit{\"a}t M{\"u}nchen,            
                   M{\"u}nchen, Germany}                                      
\centerline{$^{25}$Fachbereich Physik, University of Wuppertal,               
                   Wuppertal, Germany}                                        
\centerline{$^{26}$Panjab University, Chandigarh, India}                      
\centerline{$^{27}$Delhi University, Delhi, India}                            
\centerline{$^{28}$Tata Institute of Fundamental Research, Mumbai, India}     
\centerline{$^{29}$University College Dublin, Dublin, Ireland}                
\centerline{$^{30}$Korea Detector Laboratory, Korea University,               
                   Seoul, Korea}                                              
\centerline{$^{31}$CINVESTAV, Mexico City, Mexico}                            
\centerline{$^{32}$FOM-Institute NIKHEF and University of                     
                  Amsterdam/NIKHEF, Amsterdam, The Netherlands}               
\centerline{$^{33}$University of Nijmegen/NIKHEF, Nijmegen, The               
                  Netherlands}                                                
\centerline{$^{34}$Joint Institute for Nuclear Research, Dubna, Russia}       
\centerline{$^{35}$Institute for Theoretical and Experimental Physics,        
                  Moscow, Russia}                                             
\centerline{$^{36}$Moscow State University, Moscow, Russia}                   
\centerline{$^{37}$Institute for High Energy Physics, Protvino, Russia}       
\centerline{$^{38}$Petersburg Nuclear Physics Institute,                      
                   St. Petersburg, Russia}                                    
\centerline{$^{39}$Lund University, Lund, Sweden, Royal Institute of          
                   Technology and Stockholm University, Stockholm,            
                   Sweden, and}                                               
\centerline{Uppsala University, Uppsala, Sweden}                              
\centerline{$^{40}$Lancaster University, Lancaster, United Kingdom}           
\centerline{$^{41}$Imperial College, London, United Kingdom}                  
\centerline{$^{42}$University of Manchester, Manchester, United Kingdom}      
\centerline{$^{43}$University of Arizona, Tucson, Arizona 85721, USA}         
\centerline{$^{44}$Lawrence Berkeley National Laboratory and University of    
                  California, Berkeley, California 94720, USA}                
\centerline{$^{45}$California State University, Fresno, California 93740, USA}
\centerline{$^{46}$University of California, Riverside, California 92521, USA}
\centerline{$^{47}$Florida State University, Tallahassee, Florida 32306, USA} 
\centerline{$^{48}$Fermi National Accelerator Laboratory, Batavia,            
                   Illinois 60510, USA}                                       
\centerline{$^{49}$University of Illinois at Chicago, Chicago,                
                   Illinois 60607, USA}                                       
\centerline{$^{50}$Northern Illinois University, DeKalb, Illinois 60115, USA} 
\centerline{$^{51}$Northwestern University, Evanston, Illinois 60208, USA}    
\centerline{$^{52}$Indiana University, Bloomington, Indiana 47405, USA}       
\centerline{$^{53}$University of Notre Dame, Notre Dame, Indiana 46556, USA}  
\centerline{$^{54}$Iowa State University, Ames, Iowa 50011, USA}              
\centerline{$^{55}$University of Kansas, Lawrence, Kansas 66045, USA}         
\centerline{$^{56}$Kansas State University, Manhattan, Kansas 66506, USA}     
\centerline{$^{57}$Louisiana Tech University, Ruston, Louisiana 71272, USA}   
\centerline{$^{58}$University of Maryland, College Park, Maryland 20742, USA} 
\centerline{$^{59}$Boston University, Boston, Massachusetts 02215, USA}       
\centerline{$^{60}$Northeastern University, Boston, Massachusetts 02115, USA} 
\centerline{$^{61}$University of Michigan, Ann Arbor, Michigan 48109, USA}    
\centerline{$^{62}$Michigan State University, East Lansing, Michigan 48824,   
                   USA}                                                       
\centerline{$^{63}$University of Mississippi, University, Mississippi 38677,  
                   USA}                                                       
\centerline{$^{64}$University of Nebraska, Lincoln, Nebraska 68588, USA}      
\centerline{$^{65}$Princeton University, Princeton, New Jersey 08544, USA}    
\centerline{$^{66}$Columbia University, New York, New York 10027, USA}        
\centerline{$^{67}$University of Rochester, Rochester, New York 14627, USA}   
\centerline{$^{68}$State University of New York, Stony Brook,                 
                   New York 11794, USA}                                       
\centerline{$^{69}$Brookhaven National Laboratory, Upton, New York 11973, USA}
\centerline{$^{70}$Langston University, Langston, Oklahoma 73050, USA}        
\centerline{$^{71}$University of Oklahoma, Norman, Oklahoma 73019, USA}       
\centerline{$^{72}$Brown University, Providence, Rhode Island 02912, USA}     
\centerline{$^{73}$University of Texas, Arlington, Texas 76019, USA}          
\centerline{$^{74}$Southern Methodist University, Dallas, Texas 75275, USA}   
\centerline{$^{75}$Rice University, Houston, Texas 77005, USA}                
\centerline{$^{76}$University of Virginia, Charlottesville, Virginia 22901,   
                   USA}                                                       
\centerline{$^{77}$University of Washington, Seattle, Washington 98195, USA}  
}                                                                             
\date{\today}


\begin{abstract}
We report on a search for pair production of 
first-generation scalar leptoquarks ($LQ$)
in $p \bar{p}$ collisions at $\sqrt{s}$=1.96 TeV
using an integrated luminosity of 252 pb$^{-1}$
collected at the Fermilab Tevatron collider by the D\O\ detector.
We observe no evidence for $LQ$ production
in the topologies arising from $LQ \overline{LQ} \rightarrow eqeq$ and
$LQ \overline{LQ} \rightarrow eq \nu q$,
and derive 95$\%$ C.L. lower limits on the LQ mass 
as a function of $\beta$,
where $\beta$ is the branching fraction for $LQ \rightarrow eq$.
The limits are 241 and 218 GeV/$c^{2}$ for $\beta$=1 and 0.5, respectively.
These results are combined with those obtained by D\O\ at $\sqrt{s}$=1.8 TeV,
which increases these LQ mass limits to 256 and 234 GeV/$c^{2}$.

\end{abstract}


\maketitle

\newpage

Several extensions of the 
standard model (SM) include leptoquarks (LQ) which carry color,
fractional electric charge, and
both lepton ($l$) and quark ($q$) quantum numbers and would 
decay into a lepton and a quark~\cite{lqinfo}.
The H1 and ZEUS experiments at the $e^{\pm} p$ collider HERA
at DESY published~\cite{hera} lower limits on the
mass of a first-generation LQ that depend on the
unknown leptoquark-$l$-$q$ Yukawa coupling $\lambda$.
At the CERN LEP collider, pair production of leptoquarks
could occur in $e^{+}e^{-}$ collisions via a virtual  
$\gamma$ or $Z$ boson in the $s$-channel.
At the Fermilab Tevatron collider, leptoquarks would be pair produced
dominantly through $q \bar{q}$ annihilation (for $M_{\rm{LQ}}>$ 100 GeV/$c^{2}$)
and gluon fusion. Such pair production mechanisms are independent
of the coupling $\lambda$. Experiments at the LEP collider~\cite{lep} and
at the Fermilab Tevatron collider~\cite{run1,cdf,cdf2} set lower 
limits on the masses of leptoquarks.
In this Letter, we present a search for first-generation scalar leptoquark 
pairs produced in
$p \bar{p}$ collisions at $\sqrt{s}$=1.96 TeV  
for two cases: when both leptoquarks decay to an electron and a quark
with a branching fraction (Br) $\beta^{2}$, 
where $\beta$ is the leptoquark branching fraction into an electron and a quark,
and when one of the leptoquarks decays to an electron and a quark and 
the other to a neutrino and a quark with Br = $2\beta(1-\beta)$.
The final states consist of two electrons and 
two jets ($eejj$) or of an
electron, two jets, and missing transverse energy corresponding to the neutrino
which escapes detection ($e \nu jj$). 

The D\O\ detector~\cite{run2det} comprises three main elements. 
A magnetic central-tracking system, which consists of a silicon 
microstrip tracker and a central fiber tracker, 
is located within a 2~T superconducting solenoidal magnet. 
Three liquid-argon/uranium calorimeters, a central section (CC) covering 
pseudorapidities $\eta$~\cite{eta} with $|\eta|$ up to $\approx 1$ 
and two end calorimeters (EC) 
extending coverage to $|\eta|\approx 4$~\cite{run1det}, are housed in separate cryostats.
Scintillators between the CC and EC cryostats provide sampling of developing showers
for 1.1 $< |\eta| <$ 1.4.
A muon system is located outside the calorimeters.

The data used in this analysis were collected from April 2002 to March 2004.
The integrated luminosity for this data sample is 252 $\pm$ 16 pb$^{-1}$.
Events were required to pass at least one of a set of
electron triggers based on the requirement of one electromagnetic trigger
tower to be above threshold and on shower shape conditions.
The efficiencies of the trigger combinations
used in the $eejj$ and $e \nu jj$ analyses have been measured using data.
They are $\sim$ 100\% for two electrons  
of transverse energy ($E_T^{\rm{EM}}$) above 25 GeV, 
and for one electron above 40 GeV.
The small loss of events due to the trigger inefficiencies for $E_T^{\rm{EM}}$ below 40 GeV
is taken into account using proper weighting for Monte Carlo (MC) events.

Electrons are reconstructed as calorimeter electromagnetic clusters 
which match a track in the central-tracking system.
Electromagnetic (EM) clusters are identified by the characteristics of their 
energy deposition in the calorimeter. Cuts 
are applied on the fraction of the energy in the electromagnetic calorimeter and the isolation
of the cluster in the calorimeter.
EM clusters are marked as tight when they satisfy a shower shape condition
and loose otherwise.
Jets are reconstructed using the iterative, midpoint cone algorithm~\cite{jet}
with a cone size of 0.5.
The energy measurement of the jets has been calibrated
as a function of the jet transverse energy and $\eta$
by balancing energy in photon plus jet events.
The missing transverse energy ($\etmiss$) is calculated as the vector sum
of the transverse energies in the calorimeter cells, removing 
contributions from detector noise.

For both channels, the background arising from multijet 
events is determined from a
sample of data events (QCD sample) that satisfy the main 
cuts used in the analysis except that
each EM cluster is loose instead of tight. 
A QCD normalization factor is extracted
for this sample in a part of the phase space
where the LQ contribution is expected to be negligible. 
The QCD sample normalized by this factor is used
to derive the multijet contribution in the relevant part of the phase space.
To evaluate the $Z$ boson/Drell-Yan ($Z$/DY) and the $W$ boson background contributions, 
samples of MC events generated with {\sc alpgen}~\cite{alpgen} 
or {\sc pythia}~\cite{pythia} were used.
Samples of {\sc pythia}  $t \bar{t}$ events 
($m_{t} = 175$ GeV/$c^{2}$) were used to calculate the top quark background. 
$LQ \overline{LQ} \rightarrow eejj$ and  $LQ \overline{LQ} \rightarrow e \nu jj$
MC samples were generated using {\sc pythia} for LQ masses 
from 120 to 280 GeV/$c^{2}$ in steps of 20 GeV/$c^{2}$. 
All MC events were processed using a full simulation of the detector
based on GEANT~\cite{geant} and the
complete event reconstruction. The efficiencies of the various cuts, measured
using the data, were taken into account using proper weightings of the MC events.

\begin{figure}
\begin{tabular}{cc}
\includegraphics[scale=0.21]{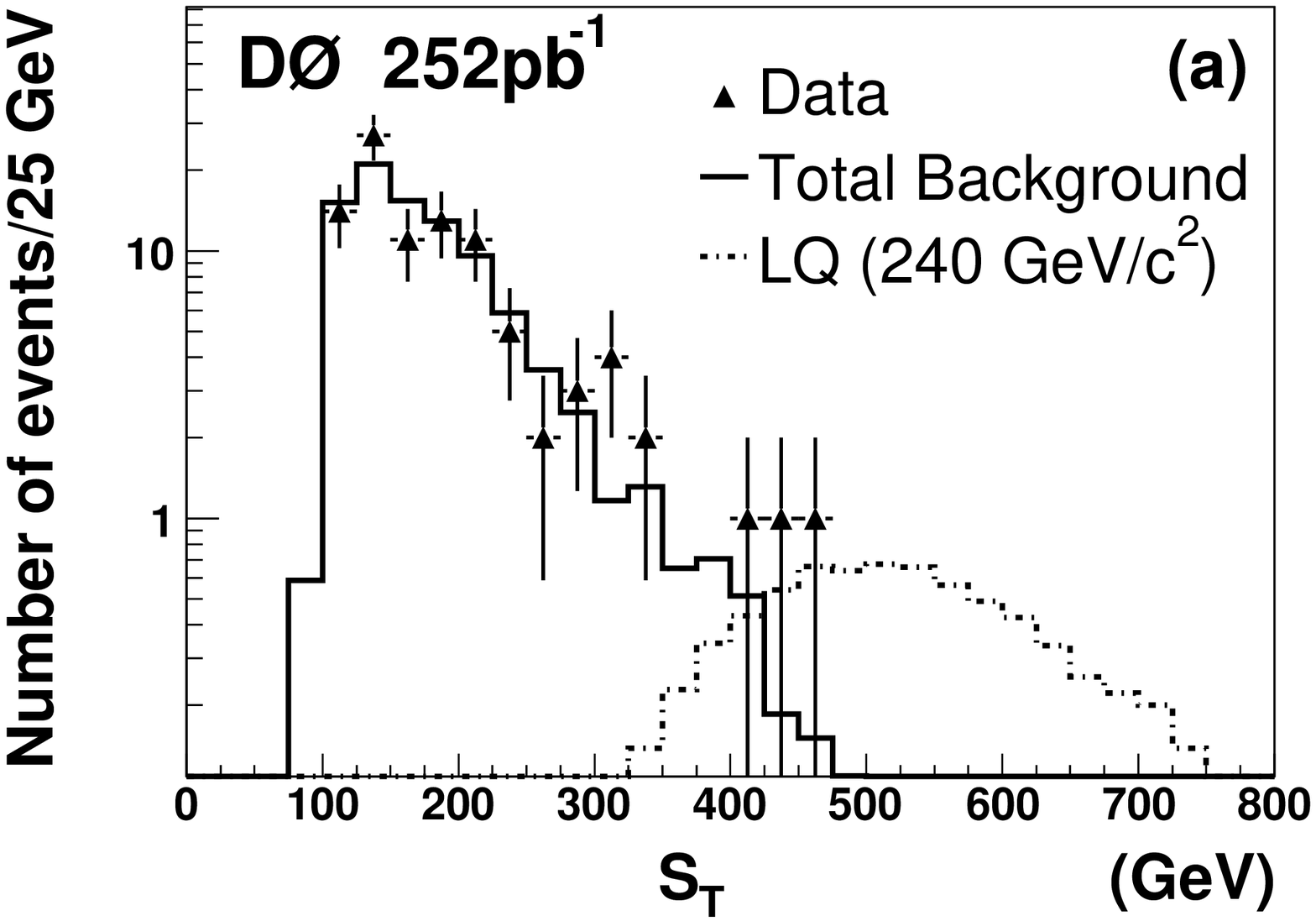} &
\includegraphics[scale=0.21]{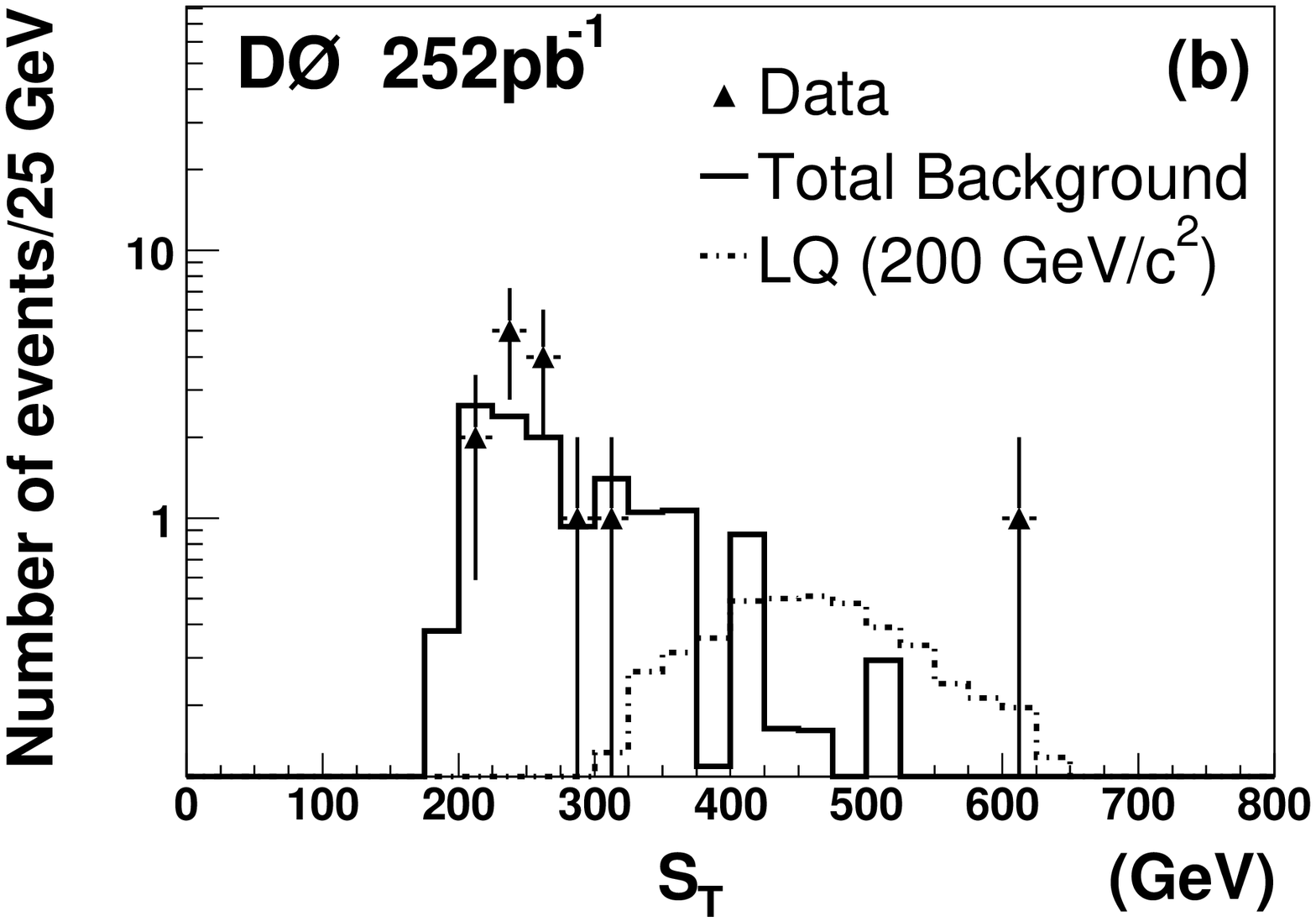}
\end{tabular}
\caption{\label{fig:st}The $S_{T}$ distributions for the $eejj$ events (a) and 
$e \nu jj$ events (b) from data (triangles)
compared to the SM background (solid histograms). 
The dot-dashed histograms are the expected distributions for a 240 GeV/$c^{2}$ LQ signal
(a) and for a 200 GeV/$c^{2}$ LQ signal (b).}
\end{figure}

The $eejj$ analysis requires two tight EM clusters
with $E_T^{\rm{EM}}$~$>$~25 GeV and at least two jets
with $E_{T}$ $>$ 20 GeV within $|\eta|<$ 2.4.
At least one of the EM clusters should spatially match
an isolated track and at least one should be in the CC
fiducial region. 
The major SM background sources that mimic the $eejj$ decay 
of a LQ pair are multijet events (where two of the jets 
are misidentified as EM objects), 
$Z$/DY production, and top quark pair production. 
To suppress background from $Z$ boson production, events with a di-electron
mass ($M_{\rm{2EM}}$) compatible with the $Z$ boson mass 
(80 GeV/$c^{2}$ $<$ $M_{\rm{2EM}}$ $<$ 102 GeV/$c^{2}$) are rejected.
Finally $S_{T} >$ 450 GeV is also required, where
$S_{T}$ is the scalar sum of the transverse energies of the two electrons 
and the two leading jets.
In Fig.~\ref{fig:st}a, the $S_{T}$ distributions 
for data and background after applying the $Z$ boson mass cut are shown.
This choice of the cutoff has been optimized using MC signal and 
background events to get the 
best expected mass limit.
The total efficiencies for a LQ signal are summarized in Table~\ref{tab:eff}.
The multijet background is estimated from 
two samples of events with two EM clusters $E_T^{\rm{EM}}$ $>$ 15 GeV 
which have at least one matched track and no reconstructed jets.
Both EM clusters are tight in one sample and loose in the other.
The QCD normalization factor is determined by the normalization of
the $M_{\rm{2EM}}$ distributions of the two samples below 75 GeV/$c^{2}$.
The $Z$/DY and top quark contributions
are normalized to the integrated luminosity.
Table~\ref{tab:ntabl6trk} lists the number of events in the data 
and the number of expected events from SM background sources.

Systematic uncertainties on the background are
determined to be 15\% from the QCD normalization factor and 6\%
from the efficiencies  of the identification of electrons and 
jets (particle-ID). An uncertainty (26\%)
from the jet energy scale is determined by varying the correction factor on the
calorimeter response to jets by one standard deviation. 
A systematic uncertainty on the $Z$/DY background (20\%) is calculated
by taking into account the differences between the two $Z$/DY MC samples. 
On the signal, the particle-ID and the limited
statistics of the MC sample correspond to systematic 
uncertainties of 6\% and 1.2\% respectively.
Comparing acceptances for the signal samples 
generated with {\sc pythia} using different parametrizations of parton
distribution functions (PDFs) leads to an uncertainty 
of 5\%. The uncertainty due to the jet energy scale is dependent on
the LQ mass (7.3\% for a LQ mass of 240 GeV/$c^{2}$). The total uncertainty on the
efficiency is (17--9)\% in the mass range 180--280 GeV/$c^{2}$. 

\begin{table}
\caption{\label{tab:eff} Efficiencies after all cuts and 95$\%$ C.L. 
upper limits on production cross section
$\times$ branching fraction Br, as a function of $M_{\rm{LQ}}$, for the two channels.}
\begin{ruledtabular}
\begin{tabular}{crcrc}
$M_{\rm{LQ}}$(GeV/$c^2$)  & \multicolumn{2}{c}{$eejj$ } & \multicolumn{2}{c}{$e \nu jj$  } \\
\hline
 & \multicolumn{1}{l}{$\ \ \ \epsilon$($\%$)} & $\sigma\times$Br(pb)&  \multicolumn{1}{l}{$\ \
 \ \epsilon$($\%$)} & \multicolumn{1}{l}{$\ \sigma\times$Br(pb)} \\
\hline
120 & 2.2$\pm$0.5 & 0.950  & 4.6$\pm$0.5 & 0.34 \\
140 & 4.5$\pm$0.9 & 0.444  & 7.9$\pm$0.8 & 0.20 \\
160 & 8.9$\pm$1.7 & 0.223 & 11.7$\pm$1.1 & 0.14 \\
180 & 12.6$\pm$2.4 & 0.156  & 15.5$\pm$1.5 & 0.10 \\
200 & 18.5$\pm$3.0 & 0.102  & 17.8$\pm$1.7 & 0.088 \\
220 & 24.6$\pm$3.5 & 0.075 & 18.9$\pm$1.8 & 0.083 \\
240 & 30.3$\pm$3.9 & 0.060 & 20.9$\pm$1.9 & 0.075 \\
260 & 34.0$\pm$4.0 & 0.053 & 21.9$\pm$2.1 & 0.071 \\
280 & 36.0$\pm$4.0 & 0.050 & 22.7$\pm$2.1 & 0.069 \\
\end{tabular}
\end{ruledtabular}
\end{table}

\begin{figure}
\begin{tabular}{cc}
\includegraphics[scale=0.22]{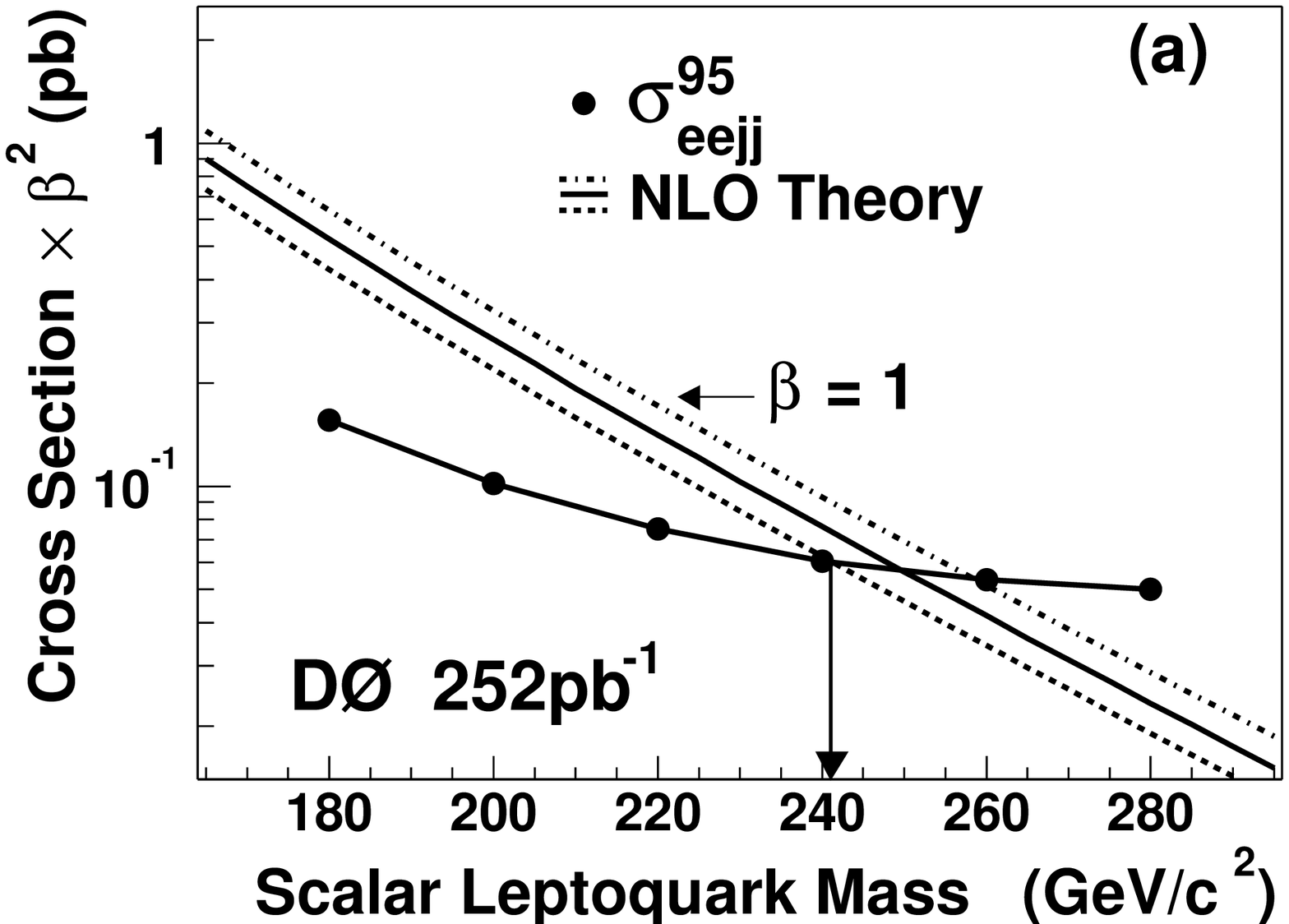}  &
\includegraphics[scale=0.22]{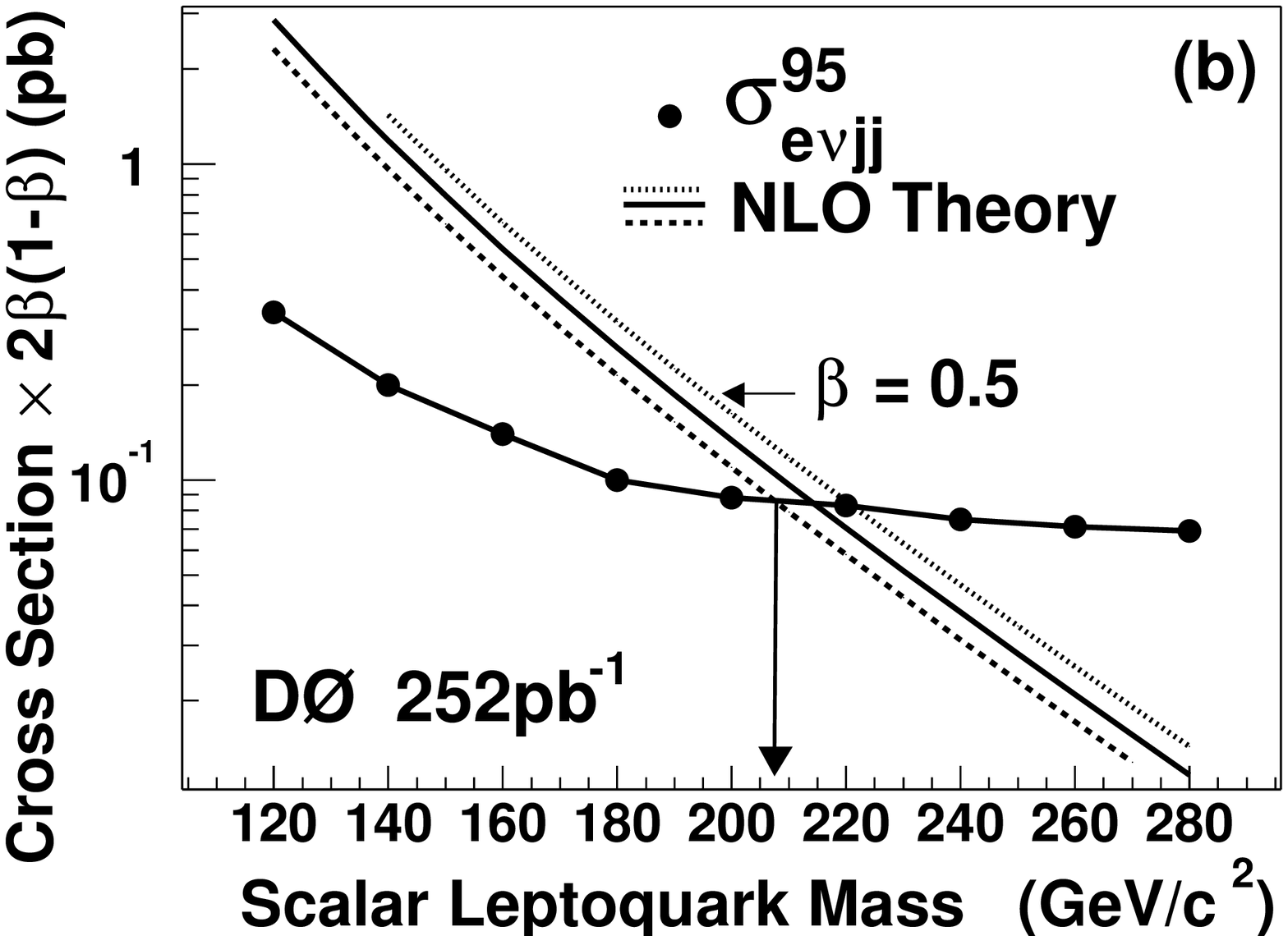}
\end{tabular}
\caption{\label{fig:limit}The $95\%$ C.L. limit on the experimental cross section times branching 
fraction as a function of LQ mass (circles) for the $eejj$ (a) 
and $e \nu jj$ (b) channels. 
The NLO theoretical cross
sections~\cite{nlo} are plotted for different values of the renormalization 
scale factor: $M_{\rm{LQ}}$ (full line), $M_{\rm{LQ}}/2 $ (dotted curve)
and $2M_{\rm{LQ}} $ (dashed curve)
taking into account the PDF uncertainties. 
A mass limit of 241 GeV/$c^2$ (a) and of 208 GeV/$c^2$ (b) 
for first-generation scalar leptoquarks
is obtained for $\beta$=1 and $\beta$=0.5, respectively.}
\end{figure}

The data are consistent with the expected SM background and no evidence for 
leptoquark production is observed in the $eejj$ channel.
Thus we can set an  upper limit at the 95\% C.L. on the LQ pair production
cross section using a Bayesian approach~\cite{bertram}.
The limits are tabulated in Table~\ref{tab:eff} and shown 
in Fig.~\ref{fig:limit}a as a function of LQ mass. To compare our experimental
results with theory, we use the next-to-leading
order (NLO) cross section for scalar leptoquark pair production
from Ref.~\cite{nlo}, with the CTEQ6 PDF~\cite{cteq}.
The theoretical uncertainties correspond to
the variation from $M_{\rm{LQ}}/2 $ to $2M_{\rm{LQ}} $ of the renormalization 
scale  $\mu$ used in the calculation 
and to the errors on the PDFs.
To set a limit on the LQ mass we compare our experimental limit to 
the theoretical cross section for $\mu$ = $2M_{\rm{LQ}}$, which is conservative as it
corresponds to the lower value of the theoretical cross section. 
The value of the theoretical cross section would increase by $\sim$ 7\% 
if the PDF errors were neglected.
A lower limit on the leptoquark mass of 241 GeV/$c^{2}$ 
is obtained for $\beta$=1.

\begin{table}
\caption{\label{tab:ntabl6trk}Number of events in data compared 
with background expectation at different stages of the $eejj$
analysis.}
\begin{ruledtabular}
\begin{tabular}{c D{,}{\,\pm\,}{-1}cc}
  &\multicolumn{1}{c}{$eejj$} & $Z$ boson veto & $S_{T}>$450 GeV \\
\hline
 Data & \multicolumn{1}{c}{467} & 95 & 1 \\
 Total background & 406 , 100 & 92$\pm$17 & 0.54$\pm$0.11 \\
 $Z$/DY + jets  & 342 , 99 & $41 \pm 11$ & 0.22$\pm$0.07 \\
 Multijet & 59 , 16 & $47 \pm 13$  & 0.27$\pm$0.08 \\
 $t \bar{t}$ production& 4.7 , 0.4 & $3.8\pm 0.3$ & 0.05$\pm$0.01\\
\end{tabular}
\end{ruledtabular}
\end{table}

The $e \nu jj$ analysis requires exactly one tight EM cluster 
($E_T^{\rm{EM}} >$ 35 GeV) in the CC fiducial region
which matches an isolated track spatially and kinematically. 
At least two jets with $E_T>25$ GeV
within $|\eta| < 2.4$ and $\etmiss >$ 30 GeV are required.
The main SM background sources which would mimic the $e \nu jj$ decay of a LQ
pair are events with multijet production  
(where a jet is reconstructed as an electron and the
$\etmiss$ comes from jet mismeasurements), 
$W$ + 2 jets events, and top quark pair production.
A veto on muons with $p_T >$ 10 GeV/$c$ 
is applied to reduce the di-lepton background from $t \bar{t}$ decays.
To suppress background from $W$ boson production, events with a transverse
mass of the electron and the missing energy 
$M_T^{e \nu}<$ 130 GeV/$c^2$ are rejected. Finally $S_T >$ 330 GeV is required,
where here $S_T$ is the sum of the transverse energies of the electron, 
the two jets, and the $\etmiss$.  The distribution of the
variable $S_T$ for the data and the total background
is shown in Fig.~\ref{fig:st}b after applying the $M_T^{\rm{e \nu}}$
cut. The choice of the cutoff has been optimized as above.
The total efficiency of these cuts for a LQ signal
is given in Table~\ref{tab:eff}.
To determine the multijet background we use a data sample
that passed all the preceeding cuts but with 
a loose EM cluster matching spatially a track.
The QCD normalization factor is determined using 
the ratio of the number of events
with $\etmiss <$ 10 GeV in this and in the search samples.
The $W$ boson background is normalized to the data
at transverse mass 60 GeV/$c^2$ $<$ $M_T^{\rm{e \nu}}$ $<$ 100 GeV/$c^2$.
The top quark background is normalized to the
integrated luminosity using the NNLO theoretical cross section.
The number of events which
survive the cuts and the number of predicted background events
are summarized in Table~\ref{tab:result}.

Systematic uncertainties associated with the QCD normalization factor (9\%) and 
$W$ boson normalization factor (5.7\%) are determined by 
the limited statistics of the samples and the choice 
of kinematical domain over which the normalization is done. 
The jet energy scale uncertainty introduces uncertainties
equal to 25\% for $W$ boson production and 8.5\% for the top-quark-pair production.
For the $W$ boson background 
an uncertainty equal to 33\% is associated to the shape of 
the $\etmiss$ distribution.
A 25\% error has been included as systematic uncertainty
on the top quark cross section. Finally,
there is an uncertainty of 3.8\% on the particle-ID acceptance. 
Three systematic uncertainties are determined on the signal acceptance:
$ 3.8 \% $ comes from the 
uncertainty on the particle-ID, $ 5 \%$ is due to the jet energy scale 
uncertainty, and $ 5.4 \% $
corresponds to the acceptance variations for different 
PDF parameterizations. 

\begin{table}
\caption {\label{tab:result} Number of events in data compared with background 
expectation at different stages of the $e \nu jj$ analysis. The values 
of the cuts are in GeV or in GeV/$c^{2}$.} 
\begin{ruledtabular}
\begin{tabular}{c D{,}{\,\pm\,}{-1} D{,}{\,\pm\,}{-1}c}
  & \multicolumn{1}{c}{$\etmiss >$ 30} & \multicolumn{1}{c}{$\  M_T^{e \nu} >$ 130}  &  $S_T >$ 330 \\
\hline
Data  & \multicolumn{1}{c}{\phantom{x} 900} & \multicolumn{1}{c}{\phantom{x} 14} &  1 \\
Total background & 902 , 211 & 13.9 , 4.4 & 3.6 $\pm$ 1.2 \\
$W$ + jets   & 811 , 211 & 10.0 , 4.4 &  2.2 $\pm$ 1.2 \\
Multijet &  76 , 7& 2.3 , 0.5 & 0.72 $\pm$ 0.28 \\
$t \bar{t}$ production & 14.7 , 2.9& 1.6 , 0.37 &  0.70 $\pm$ 0.17 \\
\end{tabular}
\end{ruledtabular}
\end{table}

As no excess of data over background is found in the $e \nu jj$ channel, 
an upper limit on the production cross section
for a first-generation scalar leptoquark is derived and 
shown in Fig.~\ref{fig:limit}b and in Table~\ref{tab:eff}. 
A comparison of these limits to theoretical calculations of the cross
section~\cite{nlo}, performed as described above, gives
a lower limit on the first-generation scalar LQ mass of 208 GeV/$c^{2}$ 
for $\beta$~=~0.5.

\begin{table}
\caption{\label{tab:sum2} 95$\%$ C.L. lower limits on the first-generation
scalar leptoquark mass (in GeV/$c^{2}$), as a function of $\beta$.
The mass limits from D\O\ ($eejj$, $e \nu jj$ and $\nu \nu jj$ combined)~\cite{run1} 
and CDF ($eejj$)~\cite{cdf}  at Run~I ($\sim $ 120 pb$^{-1}$) are also given, as well as
the limits obtained by combining the D\O\ Run I and Run II results.}
\begin{ruledtabular}
\begin{tabular}{ccccccccccc}
$\beta$ & 0.1 & 0.2 &0.3 &0.4 &0.5 & 0.6 &0.7 &0.8& 0.9& 1. \\
\hline
$eejj$ & & & & & 158 & 180 & 203 & 220 & 232 & 241\\
$e \nu jj$ & 169 & 193 & 203 & 207 & 208 & 207 & 203 & 193 & 169& \\
Comb. Run II &169 & 193 & 204 & 212 & 218 & 223& 228&232&237&241\\
D\O\ Run I &110 & & & & 204& & & & & 225\\
D\O\ Runs I $\&$ II &183 & 206 & 218 & 227 & 234 & 239& 244&248&252&256\\
CDF Run I & & & & & & & & & & 213\\
\end{tabular}
\end{ruledtabular}
\end{table}

\begin{figure}
\includegraphics[scale=0.45]{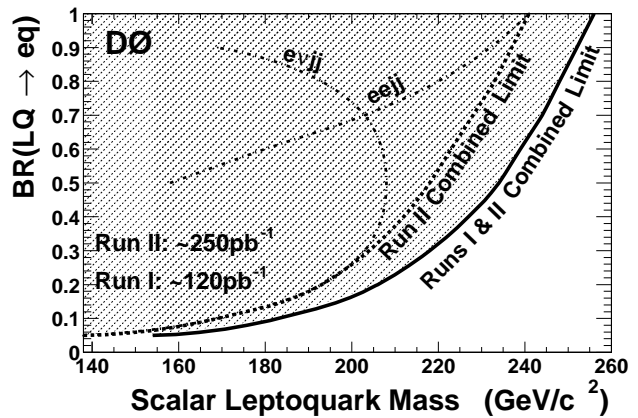} 
\caption{\label{fig:comblim}Excluded regions (shaded area) 
at the $95 \%$ C.L. in the $\beta$ versus LQ mass
plane for the production of first-generation scalar leptoquarks.}
\end{figure}

Combination of the limits obtained in the searches in the $eejj$
and $e \nu jj$ channels is done using a Bayesian likelihood
technique~\cite{paterno}, with correlated uncertainties taken into account.  
The limits on the cross sections obtained at the 95$\%$ C.L.
for the combination of the two channels and different values of $\beta$ are compared
with the NLO LQ pair production cross section~\cite{nlo} 
and lower mass limits are derived and given, as a function of $\beta$, in
Table~\ref{tab:sum2} and shown in Fig.~\ref{fig:comblim}. 
In Table~\ref{tab:sum2} are also shown the Run~I  mass limits based on 
an integrated luminosity
$\sim$~120 pb$^{-1}$ obtained by D\O\ \cite{run1},
using the three channels $eejj$, $e \nu jj$ and $\nu \nu jj$,
and CDF~\cite{cdf} ($eejj$ channel). This 
analysis sets a 95$\%$ C.L. limit on the first-generation leptoquark mass of 
$M_{\rm{LQ}} >$ 218 GeV/$c^2$ for $\beta$=0.5, 
and $M_{\rm{LQ}} >$ 241 GeV/$c^2$ for $\beta$=1.
The D\O\ Run II and Run I results are combined, using the same method, and the
results are shown in Table~\ref{tab:sum2} and in Fig.~\ref{fig:comblim}. 
The 95$\%$ C.L. limits on the first-generation leptoquark mass are
$M_{\rm{LQ}} >$~234 GeV/$c^2$ for $\beta$=0.5, and 
$M_{\rm{LQ}} >$~256 GeV/$c^2$ for $\beta$=1.

\begin{acknowledgments}
%
We thank the staffs at Fermilab and collaborating institutions, 
and acknowledge support from the 
Department of Energy and National Science Foundation (USA),  
Commissariat  \` a l'Energie Atomique and 
CNRS/Institut National de Physique Nucl\'eaire et 
de Physique des Particules (France), 
Ministry of Education and Science, Agency for Atomic 
   Energy and RF President Grants Program (Russia),
CAPES, CNPq, FAPERJ, FAPESP and FUNDUNESP (Brazil),
Departments of Atomic Energy and Science and Technology (India),
Colciencias (Colombia),
CONACyT (Mexico),
KRF (Korea),
CONICET and UBACyT (Argentina),
The Foundation for Fundamental Research on Matter (The Netherlands),
PPARC (United Kingdom),
Ministry of Education (Czech Republic),
Canada Research Chairs Program, CFI,
Natural Sciences and Engineering Research Council and 
WestGrid Project (Canada),
BMBF and DFG (Germany),
A.P.~Sloan Foundation,
Research Corporation,
Texas Advanced Research Program,
and the Alexander von Humboldt Foundation.
%
\end{acknowledgments}


\begin{thebibliography}{99}
\bibitem{eta} $\eta$ = - $\ln(\tan(\theta /2) $ where $\theta$ is the 
polar angle measured relative to the proton beam direction.
\end{thebibliography}
\end{document}